\documentclass[useAMS,usenatbib]{mn2e}

\usepackage{graphicx}
\usepackage{epstopdf}
\usepackage{subfigure}
\usepackage{lscape}
\usepackage{pdflscape}
\usepackage{txfonts}
\usepackage{times}
\usepackage{ulem}
\bibpunct{(}{)}{;}{a}{}{,}

\title[The NN Ser progenitor]{Main-sequence progenitor configurations of the NN~Ser candidate circumbinary planetary system are dynamically unstable}
\author[A.~J. Mustill et al.]{Alexander~J. Mustill$^{1}$\thanks{E-mail: alex.mustill@uam.es}, Jonathan~P. Marshall$^1$, Eva Villaver$^1$, Dimitri Veras$^2$,\newauthor
Philip J. Davis$^{3}$, Jonathan Horner$^{4,5}$ and Robert A. Wittenmyer$^{4,5}$\\
$^{1}$Departamento de F\'isica Te\'orica, Universidad Aut\'onoma de Madrid, Cantoblanco, 28049 Madrid, Spain\\
$^2$Institute of Astronomy, University of Cambridge, Madingley Road, Cambridge CB3 0HA, UK\\
$^3$Universit\'e Libre de Bruxelles, Institut d'Astronomie et d'Astrophysique, Boulevard du Triomphe, B-1050 Brussels, Belgium\\
$^{4}$School of Physics, University of New South Wales, Sydney 2052, Australia\\
$^{5}$Australian Centre for Astrobiology, University of New South Wales, Sydney 2052, Australia}

\begin{document}

\date{Accepted ---. Received ---; in original form ---}

\pagerange{\pageref{firstpage}--\pageref{lastpage}} \pubyear{2013}

\maketitle

\label{firstpage}

\begin{abstract}
Recent observations of the NN~Serpentis post-common envelope binary system have revealed eclipse timing variations that have been attributed to the presence of two Jovian-mass exoplanets. Under the assumption that these planets are real and survived from the binary's Main Sequence state, we reconstruct initial binaries that give rise to the present NN~Ser configuration and test the dynamical stability of the original system. Under standard assumptions about binary evolution, we find that survival of the planets through the entire Main Sequence lifetime is very unlikely. Hence, we conclude that the planets are not survivors from before the Common Envelope phase, implying that either they formed recently out of material ejected from the primary, or that the observed signals are of non-planetary origin.
\end{abstract}


\section{Introduction}

In recent years, eclipse timing variations have been observed in several post-common envelope binaries (PCEBs), including sub-dwarf~B binaries such as HW~Vir \citep{hwvir}, pre-Cataclysmic Variables (CVs) such as NN~Ser \citep{beuermann10}, and CVs such as HU~Aqr \citep{huaqr}. While such timing variations can be generated by angular momentum redistribution in the binary via the Applegate mechanism \citep{applegate92}, this is ruled out in many cases due to the mass of the secondary star being insufficient to generate the required magnetic field. It has been proposed that the variations are due to orbiting planets: a recent review by \cite{zs13} listed six single-planet and six two-planet candidate systems known amongst all PCEBs. Recently, \cite{pz12} used the current binary properties and two-planet solution to the HU~Aqr CV system in order to attempt to constrain the full system evolution, and determine uncertain parameters of binary evolution such as the common envelope (CE) ejection efficiency and time-scale.

However, while Keplerian fits can formally be made to the binaries' eclipse timing variations, dynamical integrations of two-planet fits thus obtained often show that the systems are violently unstable, often on time-scales of mere centuries; examples studied include HU~Aqr \citep{horner11,Hinse+12,wittenmyer12}, HW~Vir \citep{horner12b}, and NSVS~14256825 \citep*{WHM13}. Even where stability on longer time-scales is found, it may only have been demonstrated for $\sim1$\,Myr, as in the case of UZ~For \citep{Potter+11}. Such instability suggests that either the planetary parameters are radically different from the best-fit values, or that the signals are of non-planetary origin.

An exception is the pre-CV system NN~Ser, with two planet candidates whose nominal orbits are stable on 100\,Myr time-scales \citep{beuermann10,horner12a,beuermann+13}. Although the present-day stability of the planet candidates is reasonably secure, how the system evolved into its present state poses a challenge. If the system evolved from a circumbinary configuration on the Main Sequence (MS), similar to the two-planet system Kepler-47 \citep{kep_cp2}, the initial planetary configuration would have been more compact, with a wider binary, raising questions about its survival to the present day.

In this paper we seek to reconstruct the history of the NN~Ser binary, and to test the stability of the original planetary system. In Section 2 we describe our reconstruction of the initial binary configuration. In Section 3 we use the reconstructed binaries to test the dynamical stability  of the original planetary system. We address caveats in Section 4, discuss the implications of our results for NN~Ser and similar systems in Section 5, and conclude in Section 6.


\section{The binary's evolution}

\begin{figure}
  \vspace{-.5cm}
  \includegraphics[width=.5\textwidth]{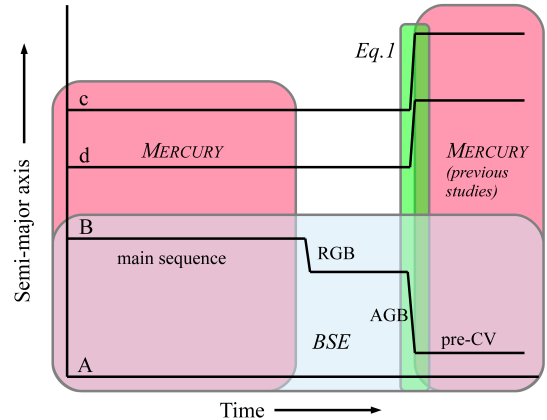}
  \caption{Schematic evolution of the orbital radii of bodies in the NN~Ser system. A and B are the primary and secondary star, c and d the planets. Italic labels state the algorithm used to study different sets of interactions at different times.}
  \label{fig:schematic}
\end{figure}

The full evolution of the NN~Ser system is complex and we must break it down into several sections (Figure~\ref{fig:schematic}). On the MS, the binary orbit would have been wider and the planets' orbits smaller. The secondary would have been too distant from the primary for its orbit to have shrunk due to tidal decay. The binary orbit may have decayed slightly due to tidal forces as the primary ascended the  red giant branch (RGB) and increased in radius, before suffering stronger tidal decay and engulfment in the envelope on the asymptotic giant branch (AGB), which resulted in the ejection of the primary's envelope and left the secondary on a very tight orbit. During the course of envelope ejection, the planets' orbits would have expanded to their present positions as a reaction to the changing gravitational potential.

Previous works have used N-body integrations to study the dynamics of the present-day system \citep{beuermann10,horner12a,beuermann+13}. In this paper we evolve binary star models forwards in time to find those that give rise to systems similar to NN~Ser. We then calculate the initial locations of the planets in the system before mass loss, and finally study the stability of the original system on the Main Sequence.

\subsection{The NN Ser binary today}

\begin{table}
\begin{center}
\begin{tabular}{lcl}
        Parameter & Value & Reference\\\hline
        $M_\mathrm{A}\mathrm{\ [M}_\odot]$  &$0.535\pm0.012$&1\\
        $M_\mathrm{B}\mathrm{\ [M}_\odot]$  &$0.111\pm0.004$&1\\
        Period [day]&$0.130\,080\,171\,41\pm0.000\,000\,000\,17$&1\\
        Separation $\mathrm{ [R}_\odot]$ &$0.934\pm0.009$& 1\\
        WD $T_\mathrm{eff}$ [K] &$57\,000\pm3\,000$&1\\
        Cooling age [yr] &$10^6$&2
\end{tabular}
\caption{Stellar parameters for the present-day NN~Ser binary. References: (1) Parsons et al. (2010); (2) Beuermann et al. (2010).}
\label{tab:NNSerStars}
\end{center}
\end{table}

NN~Ser is a pre-CV binary comprising a C/O White Dwarf (WD) of $0.535\mathrm{\,M}_\odot$ and an M dwarf of $0.111\mathrm{\,M}_\odot$ on a 0.13\,day orbit \citep{parsons10}. NN~Ser is one of the few PCEBs for which the WD properties have been determined with very high accuracy independent of any WD mass--radius relation. Furthermore, the WD mass in the NN~Ser system is larger than the peak of the distribution of He-core WDs and consistent with that of C/O WDs \citep{Rebassa-Mansergas+11}. The system is currently at an early stage of its post-CE evolution, with a cooling age of $\sim1$\,Myr. The binary properties are summarised in Table~\ref{tab:NNSerStars}.

The eclipse timing variations reported for NN~Ser suggest the presence of a planetary system. \cite{beuermann10} fit three qualitatively different solutions to the $O-C$ variability of the binary. The first, a single-planet fit, had an unacceptably large reduced chi-squared ($\chi^2_\nu=23.38$), and so two two-planet solutions with $\chi^2_\nu=0.78$ and $0.80$ were proposed. The first of these corresponds to the planets being at a 2:1 period commensurability, while the second has the planets at the 5:2 commensurability. The planet parameters for the two-planet solutions are listed in Table~\ref{tab:NNSerPls}. Stability analysis of the present-day configuration by \cite{horner12a} showed that the 2:1 solution is stable for at least $10^8$ years, while the $1\sigma$ errors on the 5:2 solution straddle the boundary between long-term stable systems, stable for $10^8$ years, and relatively unstable systems, unstable in $\sim10^5$ years. While this paper was under review, \cite{beuermann+13} provided new eclipse timings. The data remain consistent with a 2:1 solution, which can be stable for $10^8$ years, and the 5:2 solution is now ruled out. The revised parameters for the 2:1 fit are almost unchanged from \cite{beuermann10}, although the $m\sin i$ of the inner planet is somewhat smaller, at $1.74\pm0.09\mathrm{\,M_J}$ compared to $2.28\pm0.38\mathrm{\,M_J}$. In this paper, we take our system parameters from the 2:1 solution of \cite{beuermann10}. For completeness, we also briefly describe our results for the moribund 5:2 solution.


\subsection{Reconstructing the original binary}

To identify MS progenitor binaries to the current NN~Ser binary, we use the Binary Star Evolution code \citep*[BSE,][]{bse}. BSE couples evolutionary models of single stars, from \cite*{Hurley+00}, to routines governing binary interactions such as tidal forces\footnote{We correct the bug in the tidal equations described by \cite{zs13}.}, mass transfer and CE evolution. The speed of the underlying single stellar models allows the evolutionary progress of many binary systems with different parameters to be followed on computationally short time-scales. We therefore adopt an approach of modelling the evolution of large grids of models forwards in time, to identify the parameters of the progenitor binary that give rise to systems similar to NN~Ser.

In our modelling, we vary the following parameters governing the binary's initial state and evolution: the initial mass of the primary $M_\mathrm{A}$, the initial orbital period of the binary $P$ (which governs its semi-major axis $a_\mathrm{B}$), and the parameters governing the CE binding energy $\lambda$ and ejection efficiency $\alpha$. These are degenerate and only their product $\alpha\lambda$ is important. We initially fix the metallicity at $Z=0.02$, and set the binary eccentricity at zero, as this will give the most stable progenitor systems. Since NN~Ser is not yet at the CV stage, there has been no mass transfer from the secondary, and its initial mass is constrained to its current value of $0.111\mathrm{\,M}_\odot$. The possibility that the secondary may accrete material during the CE phase is discussed in Section~\ref{sec:caveat}.

We aim to find systems that give pre-CVs similar to NN~Ser after CE evolution. We require that the final masses of the stars must be within the $1\sigma$ errors given in Table~\ref{tab:NNSerStars}, while the separation must be tighter than $0.00465$\,au ($1\mathrm{\,R}_\odot$). Additionally, we place a constraint on the allowed initial binary separation. At this stage, we require simply that the initial binary can be no wider than the initial orbit of the inner planet. In Section~3 we shall test the stability of the reconstructed systems in detail, but here we note that a simple application of the results of \cite{HW99} suggests that the inner planet has to have a semi-major axis at least $1.8$ times greater than the binary semi-major axis to ensure stability on the MS.

Assuming adiabatic expansion of the planet's orbits during mass loss, the planets' initial semi-major axes are given by
\begin{equation}\label{eq:mrat}
a_\mathrm{c,d}^\mathrm{(i)}=\frac{M_\mathrm{A}^\mathrm{(f)}+M_\mathrm{B}}{M_\mathrm{A}^\mathrm{(i)}+M_\mathrm{B}}a_\mathrm{c,d}^\mathrm{(f)},
\end{equation}
where $M_\mathrm{A}^\mathrm{(i)}$ and $M_\mathrm{A}^\mathrm{(f)}$ are the primary's initial and final masses, $M_\mathrm{B}$ is the secondary's mass, and $a_\mathrm{c,d}^\mathrm{(i)}$ and $a_\mathrm{c,d}^\mathrm{(f)}$ the planets' initial and final semi-major axes. We neglect here the small contribution from the planets' masses. 
We require that these criteria be satisfied within 10\,Myr of the formation of the WD. Hence, our target space is somewhat broader than the NN~Ser system proper.

\begin{table}
\begin{center}
\begin{tabular}{lccl}
        Parameter & NN~Ser~(AB)c & NN~Ser~(AB)d & Model\\\hline
        $m\sin i/\mathrm{M_J}$&$6.91\pm0.54$& $2.28\pm0.38$ &2:1\\
        Semi-major axis/au  &$5.38\pm0.20$&$3.39\pm0.10$&2:1\\
        Eccentricity&$0$ (fixed)&$0.20\pm0.02$&2:1\\\hline
	$m\sin i/\mathrm{M_J}$&$5.92\pm0.40$& $1.60\pm0.40$ &5:2\\
        Semi-major axis/au  &$5.66\pm0.06$&$3.07\pm0.13$&5:2\\
        Eccentricity&$0$ (fixed)&$0.23\pm0.04$&5:2
\end{tabular}
\caption{Planet candidate parameters for the NN~Ser binary, from Beuermann et al.~(2010), for the 2:1 and 5:2 resonance solutions. Note that NN~Ser~(AB)d is the inner planet and NN~Ser~(AB)c is the outer planet.}
\label{tab:NNSerPls}
\end{center}
\end{table}

\subsection{Results of the binary reconstruction}

We ran a grid of BSE models to identify systems satisfying the above criteria. The initial mass of the primary ranged from $1$ to $3\mathrm{\,M}_\odot$, in steps of $0.025\mathrm{\,M}_\odot$. The parameter governing CE ejection, $\alpha\lambda$, ranged from $0.05$ to $2$ in steps of $0.05$, and then up to $4.5$ in steps of $0.5$. The initial binary period ranged from 50\,d to 250\,d in steps of 10\,d. There were thus 76\,545 binary systems in the grid calculated for binary reconstruction.

There were 369 initial binary configurations that met our criteria for the final system. These systems all had primary masses in the range $1.95-2.15\mathrm{\,M}_\odot$. The upper limit is given by assuming that the growth of the C/O core is terminated by the CE phase. The lower mass limit requires that CE evolution occur when the primary is on the AGB phase, not the RGB, setting a lower mass limit of $\sim2\mathrm{\,M}_\odot$. This is the lowest mass star that, at the assumed metallicity, does not undergo the He-flash at the RGB tip. Stars undergoing this flash achieve larger radii on the RGB. The minimum initial binary separation is $\sim0.55$\,au, which avoids tidal engulfment on the RGB. The initial separations of the successful binaries then range from $\sim0.55$\,au to $\sim1$\,au, while the CE parameter product $\alpha\lambda$ ranges from $0.5$ to $2.0$. The inner planet's orbit is at around 1\,au, meaning that all but the tightest binaries are likely ruled out by the \cite{HW99} stability estimate. Nevertheless, as this is only an approximation, and does not in any case apply to 4-body systems, we test the stability of all systems numerically in the next Section.

We subsequently generated two further grids at metallicities $Z=0.01$ and $0.03$. The youth of the NN~Ser system means that very low metallicities ($Z<<0.01$) are not likely. There was little difference compared to the nominal $Z=0.02$ case: the range of primary masses was $1.875-1.95\mathrm{\,M}_\odot$ for the low-metallicity case and $1.975-2.25\mathrm{\,M_\odot}$ for the high-metallicity case, while the initial binary orbits were respectively at $0.57-0.98$\,au and $0.55-1.05$\,au. We use only the $Z=0.02$ case as a basis for our N-body integrations.


\section{Dynamical modelling}

We now set out to determine the stability of planetary systems around our reconstructed progenitor binaries, using the hybrid symplectic integrator from the N-body code \textsc{Mercury} \citep{mercury}, with a modification to the algorithm detecting collisions with the primary described in \cite{Veras+13}. A time-step of $1/25$ of the binary's initial period is chosen, which gives maximum energy errors of $\sim10^{-5}$. We perform three sets of simulations. In the first, we fix the planets' semi-major axes $a_\mathrm{c}$ and $a_\mathrm{d}$ at a few values within the $3\sigma$ range allowed by the 2:1 solution and study the stability of the different reconstructed binary systems over a short time-scale. In the second set, we choose the most promising binary system and vary the planets' semi-major axes and masses on a finer grid consistent with the 2:1 solution, integrating these systems over the whole MS lifetime of the primary. In the third set, we again fix the binary properties but instead integrate planets consistent with the 5:2 solution.

To reconstruct the MS progenitor system we reduce the planets' semi-major axes as described by Equation~\ref{eq:mrat}. For our successful binaries, the binary initial:final mass ratio is approximately $0.3$.
 This moves the inner planet in from its current orbit of $3.39$\,au to an initial orbit of $\sim1$\,au.

The best prospects for stability assume that the planets were initially on circular orbits, and that the observed eccentricity of the inner planet is a result of subsequent dynamical excitation. On the MS we therefore set the planetary eccentricities to be initially zero. The systems are assumed to be coplanar. Our initial systems therefore consist of the binary companion at $0.5-1$\,au and two planets at $\sim1$ and $\sim1.6$\,au, all on circular orbits.

The MS lifetime of the $\sim2\mathrm{\,M}_\odot$ primaries is in excess of 1\,Gyr; hence, these systems must be stable on very long time-scales. However, due to the short period of the binary and the large number of systems we test, we initially restrict ourselves to 10\,Myr integrations to weed out the most unstable systems. For each successful progenitor binary identified in Section~2, we generate 10 realisations of the system with randomised angles, giving 3\,690 separate integrations. Although the systems are close to the 2:1 commensurability we make no effort to ensure that the resonant arguments are librating. We track when planets collide or are ejected; the ejection distance is set to 100\,au.

\begin{figure}
  \includegraphics[width=.5\textwidth]{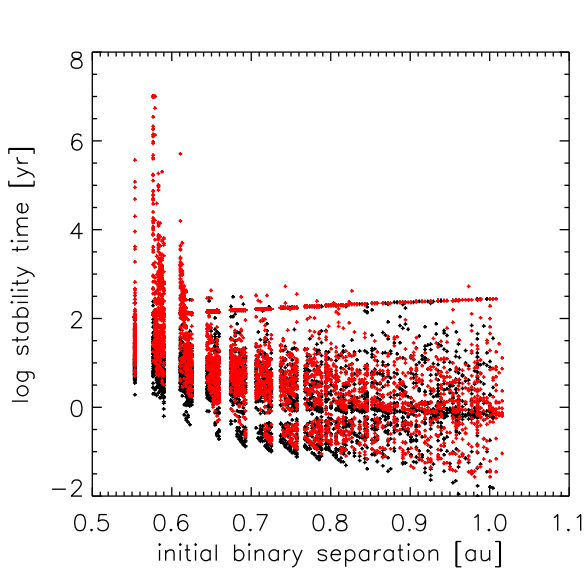}
  \caption{Timescale to instability for progenitor systems of NN~Ser, as a function of initial binary separation. Systems were integrated up to $10^7$\,yr. Black crosses show systems with the nominal planetary semi-major axes; red show systems with $a_\mathrm{d}$ increased by $2\sigma$ and $a_\mathrm{c}$ increased by $3\sigma$. The line of points at $\sim100$\,yr is an artefact of \textsc{Mercury}'s ejection-tracking algorithm.}
  \label{fig:tstab-systems}
\end{figure}

The stability lifetimes of these systems are shown as the black points in Figure~\ref{fig:tstab-systems}. All of these systems with the planets' semi-major axes set to the adiabatically contracted nominal values were unstable on very short time-scales of $<400$\,yr. We therefore tried increasing the planets' axes by $1$, $2$ and $3\sigma$ in order to find more widely-separated and therefore stable systems. The number of integrations was around 3\,700 in each set, changing slightly with the inner planet's semi-major axis (which determines the number of successful binary progenitors). The most stable configuration was with the inner planet's orbit enlarged by $2\sigma$ and the outer planet's by $3\sigma$. The stability lifetimes of these systems are shown as the red points in Figure~\ref{fig:tstab-systems}. 16 of the 3\,740 systems integrated survived for the 10\,Myr integration. These were all orbiting fairly close binaries, with $a_\mathrm{B}<0.6$\,au\footnote{While in this grid all the stable systems satisfied the \cite{HW99} criterion, we did find instances (e.g., when increasing both planets' semi-major axes by $3\sigma$) where the stable systems' inner planets orbited inside the stability boundary from \cite{HW99}. The boundary estimate fails by a few percent, within the errors quoted in \cite{HW99}.}. Hence, although the number of stable systems is small ($0.4$\%), we find the possibility that systems may survive the entire pre-CE lifetime. We pursue this further in the next section.

\subsection{Variation of planet parameters}

\begin{figure}
  \includegraphics[width=.5\textwidth]{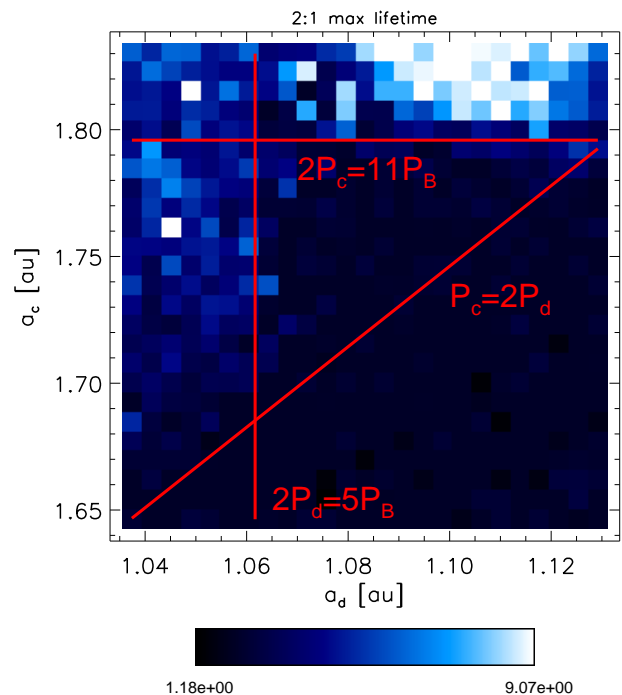}
  \caption{Stability maps for the reconstructed NN~Ser system. The semi-major axes of inner and outer planets are varied. The nominal values are those at the bottom left corner of the plots, while the axes extend to the nominal value plus $3\sigma$. Colour scale shows the maximum lifetime (in years, log scale) of systems at each combination of axes. Some important mean motion resonances are marked in red.}
  \label{fig:stable-grid}
\end{figure}

To explore the 2:1 solution in more depth, we now consider one of the binary systems that showed promise of long-term stability in the previous section. This binary initially had $M_\mathrm{A}=2.0\mathrm{\,M}_\odot$, $P=110$\,days, $a_\mathrm{B}=0.57$\,au. We ran a grid of simulations varying the planet's semi-major axes from their nominal values out to a $3\sigma$ increase\footnote{We did not consider reducing the planets' semi-major axes as compressing the system will likely make it yet more unstable.}, with 25 semi-major axis values for each planet and 10 systems with random initial angles for each combination of semi-major axes: a total of 6\,250 systems. These were integrated until they suffered a collision, close encounter, or ejection, up to a maximum time of $1.164$\,Gyr, at which the primary leaves the MS.

The vast majority of these systems are unstable on very short time-scales. The maximum lifetime of the 10 systems at each grid point are shown in Figure~\ref{fig:stable-grid}. In all, only 16 systems, $0.26$\%, survived until the end of the MS. Long-lived systems, and particularly those that survive the whole MS, are concentrated with $a_\mathrm{d}$ between $1.08$ and $1.12$\,au and $a_\mathrm{c}$ beyond $1.80$\,au. A second region of moderate stability, with two survivors, lies with $a_\mathrm{d}<1.06$\,au. Nevertheless, the low number of survivors, and the fact that these are found in a region of parameter space far from the nominal values, suggests that the evolution of the present NN~Ser planetary system from a more compact configuration is rather unlikely.

We also ran a second grid, fixing the planets' semi-major axes at their nominal values but reducing the masses by up to $3\sigma$, again on a $25\times25$ grid with 10 realisations per point. We note that the new lower mass for the inner planet \citep{beuermann+13} is covered by this grid. None of these systems lasted longer than 1500\,yr. The dominant effect of the binary, together with the relative insensitivity of analytical stability criteria to small changes in mass, mean that the planets' semi-major axes exert the dominant influence on stability.

We performed the same analysis for the 5:2 solutions, finding them even more unstable: only 7 of 6\,250 systems survived for the primary's MS lifetime when allowing $a_\mathrm{c}$ and $a_\mathrm{d}$ to vary, and the longest-lived system had a lifetime of only 500 years when allowing $m_\mathrm{c}$ and $m_\mathrm{d}$ to vary.


\section{Other evolutionary paths for the NN~Ser system}

\label{sec:caveat}

The results of the stability analysis presented above suggest that a very small fraction of potential progenitor systems for NN~Ser could have survived to the end of the primary's MS lifetime, and that these require orbital elements that are rather removed from their nominal values. While the evolution of the present-day two-planet configuration from a MS system is therefore a possibility, its likelihood seems rather low. Is there any way of mitigating these harsh conclusions, and finding some evolutionary path for the system?

\subsection{Uncertain binary/CE physics}

The most important objection is that binary evolution may permit binaries on tighter initial orbits than those we have considered. However, tight orbits require high values of $\alpha$ in order to achieve a given final outcome, and high values of $\alpha$ are not favoured either by observations of PCEBs or by theoretical considerations \citep[see][]{Zorotovic+10,DeMarco+11,DKK12,Rebassa-Mansergas+12}; in fact, these works suggest $\alpha$ values at the lower end of the range considered in this paper, with $\lambda$ typically in the range $0.1-0.4$ and $\alpha$ from $0.2-0.3$, and certainly $<1$.

Furthermore, attempts to seek closer initial binaries run up against the problem of engulfment on the RGB. This may be averted if the strength of tidal forces were much weaker than included in the BSE code. While studies of tidal circularisation in binaries suggest that tidal dissipation should be stronger than the \cite{z77} theory employed in BSE \citep{mm05}, numerical simulations of tidal dissipation of the convective zones of Solar-type MS stars suggest weaker tides \citep{penev+09}. Reducing the strength of tidal forces by a factor of 10 reduces the orbit of the tightest binary that can survive the RGB to $0.34$\,au, and test integrations show that around half of the systems tested with $a_\mathrm{B}<0.5$\,au are stable for 10\,Myr, although we have not followed their evolution on Gyr timescales.

Alternatively, engulfment on the RGB may not be a problem if the WD is in fact a He-core WD, in which case engulfment would indeed have happened on the RGB. This option does not however appear to be supported by detailed observations of the WD's spectrum \citep{parsons10}.

Another possibility is that the initial mass of the companion was smaller than its current mass, and that it accreted material during the CE phase. However, such accretion would be limited because the secondary expands as a result of the accretion, causing mass exchange to stabilise once the secondary fills its Roche lobe: \cite{HT91} showed that, at least in the case of a $1.25\mathrm{\,M}_\odot$ MS secondary, the effects of the CE phase on the secondary mass are small. With such a low mass secondary as in the NN~Ser system, accretion will be even less significant. Indeed, \cite{Maxted+06} found that the post-CE brown dwarf companion to WD 0137--349 has a spectral type consistent with a brown dwarf that has seen little accretion during the CE.

\subsection{Unpredictable dynamics}

It may also be that our assumption of adiabatic expansion of the planets' semi-major axes may be incorrect. We note that we require some form of excitation to raise the inner planet's eccentricity to its observed value---the eccentricities of planets surviving our integrations did not exceed $0.07$---and non-adiabatic orbit expansion due to the rapid loss of mass during the CE phase provides a natural source \citep{pz12}. In general, rapid mass loss may cause the orbits to expand either faster or slower than the adiabatic rate \citep{Veras+11}, and if the semi-major axis increase were lower, the planets would have initially been further from the binary and likely more stable. However, when starting from a circular orbit, rapid mass loss causes the semi-major axis to expand faster than in the adiabatic case. This is because angular momentum must still be conserved:
\begin{equation}
\sqrt{\mathcal{G}\left(M_\mathrm{A}^\mathrm{(f)}+M_\mathrm{B}\right)a_\mathrm{c,d}^\mathrm{(f)}\left(1-e_\mathrm{c,d}^{\mathrm{(f)}2}\right)}=\sqrt{\mathcal{G}\left(M_\mathrm{A}^\mathrm{(i)}+M_\mathrm{B}\right)a_\mathrm{c,d}^\mathrm{(i)}\left(1-e_\mathrm{c,d}^{\mathrm{(i)}2}\right)},
\end{equation}
where (i) and (f) label initial and final elements. To change a circular orbit to one at a final $e_\mathrm{c,d}^\mathrm{(f)}\ne0$ requires that
\begin{equation}
a_\mathrm{c,d}^\mathrm{(i)}=\frac{M_\mathrm{A}^\mathrm{(f)}+M_\mathrm{B}}{M_\mathrm{A}^\mathrm{(i)}+M_\mathrm{B}}a_\mathrm{c,d}^\mathrm{(f)}\left(1-e_\mathrm{c,d}^{\mathrm{(f)}2}\right),
\end{equation}
a rate of semi-major axis expansion faster than that given in Equation~\ref{eq:mrat}. Starting from an eccentric orbit would allow a smaller increase in $a$ but the eccentricity would compromise the stability of the initial system.


We note that we have not attempted to ensure that the original system was in an exact resonant state, which can enhance stability in two-planet systems. However, we estimate that conjunctions of the inner planet with the secondary would cause large enough perturbations to swiftly remove the two planets from resonance \citep[][equation 9.93]{md}, and it is unlikely that the system could maintain a resonant configuration throughout its MS lifetime. It is possible that the present resonant or near-resonant configuration was attained recently. This would be possible if the planets migrated during the CE ejection, but we argue below that we expect drag forces to be negligible and little migration to occur. It is also possible that the system has been recently perturbed into its present state by an internal instability or a passing star. While attempting to reproduce the outcome of such an event could only be done statistically, it is likely that any possible progenitor configuration would also run into stability problems when on the MS.


\subsection{CE ejecta and drag}

It may be possible that the planets started further away from the binary than we have reconstructed. A slight reduction of the planets' semi-major axes may also be possible through drag effects as the envelope is expelled. In order to estimate how the planets' orbits will shrink as a consequence of the density increase in the circumstellar environment associated with the ejection of the envelope, we use the same formalism as in \cite{VL09}, including both the the gravitational drag force $F_\mathrm{g}$ and the frictional force $F_\mathrm{f}$ according to the following expressions:
\begin{equation}
F_\mathrm{g}= 4\pi \frac{(G m_\mathrm{pl})^2 }{c_\mathrm{s}^2}\rho I~~,
\label{eq:fg}
\end{equation}
with $I\simeq0.5$, $G$ being the gravitational constant, $\rho$ the density encountered by the planet, $m_\mathrm{pl}$ the planet mass, and $c_\mathrm{s}$ the sound speed;
\begin{equation}
F_\mathrm{f} = \frac{1}{2} C_\mathrm{d} \rho v^2 \pi R_\mathrm{pl}^2~~,
\label{eq:ff}
\end{equation}
where $C_\mathrm{d}\simeq0.9$ is the dimensionless drag coefficient for a sphere and $R_\mathrm{pl}$ is the radius of the planet and $v$ its orbital velocity.

In order to estimate the density for the calculation of the drag terms we consider the spherically symmetric density profile created by a wind launched from the primary's surface at a velocity of $10$ and $15\mathrm{\,km\,s}^{-1}$ at a rate of $1.5\times10^{-3}\mathrm{\,M}_\odot\mathrm{\,yr}^{-1}$. These outflow velocities are typical of winds ejected at the sound speed of the envelope, while such high mass-loss rates, or still higher, must arise from the brief duration of the CE phase. Having established the density profile, we then treat the envelope as static and unchanging during the ejection, which continues for 10\,000 years. This therefore significantly overestimates the total effect of the drag force, as the force is integrated over a longer time period than can be maintained at such high mass loss rates, and also in reality the planets will only see the highest densities for a fraction of the period of envelope ejection. Even so, we find that the planets' orbits decay by only a few hundredths of an au, insufficient to move the initial locations to the more stable regions identified in Figure~\ref{fig:stable-grid}. We note that CE ejection in reality will not be isotropic and it is likely that the bulk of material is ejected in the orbital plain. Our calculations overestimate the total impulse on the planet by about a factor of 10, and so account for the expected mid-plane density enhancements \citep{RickerTaam12}.

We do acknowledge the possibility that a disc of material may remain bound to the binary after ejection of the envelope. Detailed numerical studies of CE evolution \citep[e.g.,][]{KS11} typically result in some fraction of the material of the primary's envelope's remaining bound to the binary after the rapid in-spiral phase. However, the quantity of such material, as well as its location after fall-back, is not well-constrained. If sufficient quantities remain at several au for sufficiently long time periods, it may trigger planet migration as in protoplanetary discs.

\section{Discussion}

Because survival of a two-planet system from the MS appears highly improbable, we must seek other explanations for the eclipse timing variations.

Firstly, it could be the case that there is only one planet in the system. Although the one-planet fit of \cite{beuermann10} has a much greater reduced $\chi^2$ than either of the two-planet fits, in its original configuration the planet would have been at $\sim2$\,au, comfortably distant from the binary companion to survive the primary's MS evolution.

Secondly, perhaps the present planetary system did not evolve directly from any MS planetary system, instead representing the outcome of second-generation planet formation from debris from the ejected stellar envelope. We note that the short time-scale for envelope ejection may however leave little time for such planet formation. Furthermore, the large mass of the planets implies that the disc they formed from was rather massive.

Second-generation planet formation has received scant theoretical attention. As discussed above, models of CE ejection provide the possibility that discs may form. Very large (up to 1\,000s of au) circumbinary discs have been observed orbiting some post-AGB stars such as the Red Rectangle \citep{Bujarrabal+05}; these discs may arise from the effects of jets on material in the AGB wind \citep{AkashiSoker08}. However,it is not known how the properties of the disc scale with different binary parameters and mass loss rates, nor is it known whether planets can actually be created in the disc should one form. Other scenarios for second generation planet formation, such as AGB wind accretion in wide binaries \citep{PeretsKenyon13} and supernova fallback discs \citep{Hansen+09}, are less relevant to a post-CE situation. Second generation planet formation in the NN~Ser system remains a speculation.

Lastly, the eclipse timing variations may not be of planetary origin at all, but instead reflect some as-yet unexplained phenomenon. When NN~Ser is considered in the context of other post-CE binaries with planet candidates, which have often been shown to be dynamically implausible, a non-planetary origin for the timing variations perhaps becomes the more likely explanation. If the variations cannot be planetary in other systems, it is only natural to suppose that the same process is at work in all systems, and that in this one case it so conspires that the timings can be interpreted as a stable planetary system. Proof of the planetary origin of the period variations would come from detection of the planets' mutual gravitational interactions, although these will take several decades to become apparent \citep{beuermann+13}.

While the number of post-CE binaries with eclipse timing variations attributed to planets is growing, the planetary origin of these is coming increasingly under attack. \cite{horner11,horner12b}, \cite{Hinse+12}, \cite{wittenmyer12}  and Horner et al. (submitted) show that many of the proposed two-planet systems are unstable on very short time-scales, convincingly showing that the origin of the variations cannot be attributed to planetary systems with the architectures suggested by Keplerian fitting. NN~Ser, the system studied here, is the only two-planet system that has been proved to be long-term stable in its currently observed state.

\cite{zs13} used a statistical argument to refute the variations' being attributable to planetary systems that have evolved from the MS: in nearly all systems where the eclipse timing precision is sufficient to detect the presence of planetary mass companions, such companions have been found, but the occurrence rate of giant planets orbiting the progenitor MS binaries is much lower. They conclude that either second-generation planet formation is highly efficient during the envelope ejection, or that the eclipse timings are of a non-planetary origin. Our detailed investigation of the NN~Ser system supports this conclusion: it is highly unlikely that the planetary system evolved from a MS state. If the planets are real, they must be of second-generation origin. Perhaps the more likely alternative, however, is that they do not exist.

\section{Conclusion}

There are three explanations for the eclipse timing variations observed in NN~Ser: they may be due to perturbations from first-generation planets that survived the whole evolution of the binary; they may be due to second-generation planets that formed from remnant material after ejection of the primary's envelope; or they may be due to some process intrinsic to the binary. Based on the calculations in this work, the first possibility seems unlikely.

We have reconstructed possible MS progenitors of the NN~Ser pre-CV binary, and shown that almost all initial configurations of the planetary system are unstable on very short time-scales. We acknowledge that a full exploration of all values of all parameters is not possible, and hence we may have missed some possible solution. However, perhaps the biggest uncertainty in modelling the evolution of close binaries, and that which has the greatest effect on the evolution, lies in the values of the CE parameters, with which we have dealt. Furthermore, we note that we have not dealt with several aspects of the transition from the MS to the pre-CV binary: notably, the effects of the in-spiral of the secondary during the RGB, AGB and CE phases, which may cause eccentricity excitation through mean motion resonance sweeping, and the eccentricity excitation due to non-adiabatic mass loss. While some of the latter may be required to pump the eccentricity of the planet to its present value, too much would easily destabilise the system. Hence, the prospects for finding a consistent evolution from the MS through to the present day are even more dire than we have presented. We are left with three choices: either there is only a single planet, although the solution is a poor fit to the data; or there are two planets which recently formed out of matter from the ejected envelope; or the eclipse timing variations are of non-planetary origin. Considering that in other similar systems a planetary origin has been shown to be highly unlikely, a non-planetary origin is also our preferred scenario for NN~Ser.

\section*{Acknowledgments}

JPM is supported by Spanish grant AYA 2011/26202. AJM and EV are supported by Spanish grant AYA 2010/20630. EV also acknowledges the support of the Marie Curie grant FP7-People-RG268111. We thank two anonymous referees for helping us to improve and clarify the paper.

\bibliographystyle{mn2e-long}
\bibliography{nn_ser_evol}

\bsp

\label{lastpage}

\end{document}